\documentclass[12pt,preprint]{aastex}
\newcommand{\um}{\,$\mu$m}
\newcommand{\sst}{{\it Spitzer}}
\newcommand{\hst}{{\it Hubble}}

\def\deg{\ifmmode {^{\circ}}\else {$^\circ$}\fi}

\slugcomment{{\it Astrophysical Journal}, in press}
\shorttitle{A J-dropout at z$\sim$2 ?}
\shortauthors{Chary et al.}

\begin{document}

\title{HUDF-JD2: Mid-infrared Evidence for a $z\sim2$ Luminous Infrared Galaxy}

\author{Ranga-Ram Chary\altaffilmark{1}, Harry I. Teplitz\altaffilmark{1}, Mark E. Dickinson\altaffilmark{2}, 
David C. Koo\altaffilmark{3}, Emeric Le Floc'h\altaffilmark{4,7}, 
Delphine Marcillac\altaffilmark{5}, Casey Papovich\altaffilmark{5,7}, Daniel Stern\altaffilmark{6}}
\altaffiltext{1}{{\it Spitzer} Science Center, California Institute of 
Technology, Pasadena, CA 91125; {\tt rchary@caltech.edu}}
\altaffiltext{2}{NOAO, 950 N. Cherry St., Tucson, AZ 85719}
\altaffiltext{3}{UCO/Lick Observatories, Santa Cruz, CA 95064}
\altaffiltext{4}{Institute for Astronomy, Honolulu, HI 96822}
\altaffiltext{5}{Steward Observatory, University of Arizona, Tucson, AZ 85721}
\altaffiltext{6}{Jet Propulsion Laboratory, Pasadena, CA 91109}
\altaffiltext{7}{\sst\ Fellow}

\begin{abstract}
The \hst\ Ultra Deep Field source JD2 presented in Mobasher et al. (2005)
is an unusual galaxy that is very faint at all wavelengths shortward 
of 1.1$\mu$m. Photometric redshift fits to data at 0.4 to 8\,\um\ yield a significant
probability that it is an extremely massive galaxy at $z\sim6.5$. In this
paper we present new photometry at 16\,$\mu$m and 22\,$\mu$m from \sst\
Infrared Spectrograph (IRS)
peak-up imaging of the Great Observatories Origins Deep Survey (GOODS) fields. 
We find that the spectral
energy distribution shows a factor of $\sim$4 rise in flux density between
the 16\,$\mu$m and 22\,$\mu$m bandpass which is most likely
due to the entrance of 
polycyclic aromatic hydrocarbon emission features
into the 22\um\ and 24$\mu$m passbands. The flux ratio between
these bandpasses can be best fit by a $z\sim1.7$ luminous infrared
galaxy with a bolometric luminosity of (2$-$6)$\times$10$^{11}$~L$_{\sun}$ corresponding to 
a star-formation rate of 80~M$_{\sun}$~yr$^{-1}$.
The predicted flux density values at other longer wavelengths are 
below the detection limits of current instrumentation but
such sources could potentially be detected in lensed submillimeter surveys.
Re-evaluation of the optical/near-infrared photometry continues to favor $z>6$
photometric redshift solutions, but we argue that the consistency of the
multiwavelength parameters of this galaxy with other dusty starbursts 
favor the $z\sim2$ mid-infrared photometric redshift.
The data presented here provide evidence that optically undetected
near-infrared sources which
are detected at 24\,$\mu$m are most likely dusty, starburst galaxies
at a redshift of $z\sim2$ with stellar masses $>$10$^{10}$~M$_{\sun}$. 

\end{abstract}


\section{Introduction}

JD2 is an interesting object in the Hubble Ultradeep Field \citep[UDF;][]{RIT, Beckwith}.
Its non-detection shortward of the 1.1$\mu$m bandpass,
even in ultradeep Advanced Camera for Surveys (ACS) UDF optical data which 
are sensitive down to $\sim$29 AB mag, has been interpreted
as being due to the redshifted Lyman break. In addition, it displays a break
in its broad band spectral energy distribution (SED) between 2.2$\mu$m and 3.6$\mu$m  which has been interpreted as
the redshifted Balmer break. 
\citet{Mobasher} argue that it may be a very massive (6$\times$10$^{11}$~M$_{\sun}$) galaxy with very little
ongoing star-formation at $z\sim6.5$.
The large stellar mass of this galaxy derived from fitting population synthesis models to the multiband photometry,
and high source density implied by the presence of one such object in the small area subtended by
the Near-infrared Camera and Multi-object Spectrograph (NICMOS) UDF, suggest 
that UV-faint objects, if indeed at such high redshifts,
contribute as much to the stellar mass density at $z\sim6.5$ as rest-frame ultraviolet-bright Lyman break galaxies. 
Since the age of the stellar population in JD2 is thought
to be $\sim$600 Myr, galaxies like JD2 might harbor remnants of
the first epoch of star-formation, and play an important role in the
reionization of the intergalactic medium (IGM). Such an object, if truly at high redshift,
would challenge current models of galaxy formation, which do not predict a large
number density of galaxies more massive than 10$^{11}$~M$_{\sun}$ at $z>6$ \citep{Dave:06}.

The photometric redshift solutions in \citet{Mobasher} favor a $z\approx6.5$ solution but provide
a 15\% probability that the source is at $z<5$, with the most likely alternative solution implying a
dusty galaxy at $z\approx2.5$.
There are a few inconsistencies in the high-redshift
interpretation of JD2. The optical/near-infrared SED is best fit by a template
without significant extinction or star-formation. However, the object is detected in the Great Observatories Origins Deep
Survey (GOODS) \sst\ Multiband Imaging Photometer and Spectrometer (MIPS) 
24$\mu$m survey with a flux density of
51.4$\pm$4$\mu$Jy. Since the 24$\mu$m flux would be dominated by redshifted
hot dust and polycyclic aromatic hydrocarbon emission, it indicates the presence of dust within the galaxy.
The age-extinction degeneracy is very well known for red galaxies. Most recently, \citet{Stern:06} have shown that
the optical/near-infrared photometry of dusty, extremely red objects (EROs) is almost identical to that
of evolved, passive EROs, and that only the detection of reprocessed emission at mid- and 
far-infrared wavelengths can break
the degeneracy. Nevertheless,
Mobasher et al. (2006) suggest that the 24$\mu$m
emission could be explained by the presence of an obscured
active nucleus. While this is possible, at $z\sim6.5$,
the {\it Chandra} 2$-$8 keV band traces 15$-$60 keV emission, energies which are relatively immune to absorption. 
The $Chandra$ detection limit corresponds to
L$_{\rm X}$ of 3$\times$10$^{43}$~erg~s$^{-1}$ at $z=6.5$ while the mid-infrared detection
implies $\nu$L$_{\nu}$ at rest 3\um\ of 3$\times$10$^{45}$~erg~s$^{-1}$. The 10\um\ 
to X-ray luminosity ratios of local Seyfert nuclei are $\sim$3 \citep[e.g.][]{Krabbe}, and the flat
spectrum of Seyferts in $\nu L_{\nu}$, implies that
either large (N$_{\rm H}>>10^{24}$\,cm$^{-2}$) column densities of neutral gas must be obscuring the hard X-ray emission
or the active galactic nucleus (AGN) is an unusually low luminosity X-ray source. 

In a recent paper, \citet{Dunlop}
question the optical limits in the $B$, $V$, $i$ and $z$ passbands
adopted by \citet{Mobasher} and demonstrate that the $\chi^{2}$ values in the
redshift fits are skewed in the \citet{Mobasher} analysis
by adopting optical flux limits which are too stringent. 
By refitting the source with revised optical limits, they conclude 
the source to be at $z\sim2.2$ rather than at $z\sim6.5$.
We also note that JD2 has been visited in work by \citet{Yan:04} and \citet{Chen} prior to the
\citet{Mobasher} analysis,
and had been classified as a $z\sim3$ ERO.

In this paper, we present new 16\um\ and 22$\mu$m photometry of this source from \sst\ 
Infrared Spectrograph (IRS) peak-up imaging of the
GOODS fields \citep{Teplitz:07}. We fit the photometry at these two wavelengths and the published
24\um\ data with a variety
of dust spectral energy distributions and demonstrate that the mid-infrared photometry is most
consistent with a redshift of $z\sim1.7$ for JD2. We assess the accuracy
of the optical/near-infrared photometry and refit the data with this redshift constraint
to determine physical properties of the galaxy. We conclude that it is a luminous infrared galaxy
with L$_{\rm IR}\sim10^{11.7}$~L$_{\sun}$ and not a $z>6$ galaxy with a dust obscured AGN. Throughout
this paper, we adopt a $\Omega_{M}=0.27$, $\Omega_{\Lambda}=0.73$, H$_{0}$=71 km~s$^{-1}$~Mpc$^{-1}$
cosmology.

\section{IRS Peak-Up Imaging of GOODS-S }

The 16$\mu$m and 22$\mu$m observations were carried out using the peak-up imaging capability of the Infrared Spectrograph
(IRS) instrument \citep{Houck:04} on the \sst\ {\it Space Telescope}. The field of view of the peak-up imaging
camera is 54$\arcsec\times$80$\arcsec$. The central $\sim$130~arcmin$^{2}$ of the
GOODS-S field was observed with 60s frame times. Most of the area within the GOODS fields had four dithers
per position while the UDF area had 32 dithers per position. Each frame was distortion corrected and 
background subtracted. The dithered frames were combined 
using ``drizzle'' \citep{Fruchter}. Due to the limited number of dithers per position, point kernel drizzling, 
which would have minimized the correlated noise in the mosaics was not possible. The final mosaics
have a plate scale of 0.9$\arcsec$ per pixel. The exposure time in the final mosaics is
$\sim$4 min~pix$^{-1}$ over most of the GOODS area while the UDF area was observed with 32 min pix$^{-1}$. 
JD2, although in the UDF region, seredipitously lies in a part of the coverage map where the exposure time is 68 min,
twice that of the nominal UDF. As a result, the statistical uncertainty on its flux is lower than for the average
UDF source of the same brightness.

The spatial resolution of $Spitzer$ at 16$\mu$m and 22$\mu$m 
is 4.1$\arcsec$ and 5.2$\arcsec$ full width at half maximum (FWHM), respectively. At this 
resolution, the majority of sources, including JD2,
are point sources. This makes it possible to use prior positions from higher resolution
data, such as the GOODS 3.6$-$8\um\ imaging \citep[][Dickinson et al., in prep.]{MED03} and apply point-source fitting techniques
to measure the flux density of the source. The technique has successfully been used for cataloging 
24$\mu$m sources from the GOODS/MIPS imaging survey.  The 
results are consistent with aperture photometry, with appropriate aperture corrections, for
isolated sources and alleviate the contamination to the photometry
from the wings of the point spread function for sources which have nearby companions.
Postage stamp image cutouts of JD2 are shown in Figure 1.

The 16$\mu$m and 22$\mu$m flux densities of the source are 13.5$\pm$3.5\,$\mu$Jy and 56$\pm$12\,$\mu$Jy respectively.
In comparison, the 24\,$\mu$m flux density of the source from the GOODS imaging of the field is 51$\pm$4\,$\mu$Jy.
HUDF-JD2 is separated by 7$\arcsec$ from a brighter mid-infrared
source to the south east which is a $z=0.457$ spiral galaxy (Stern et al., in prep.). 
In order to estimate the systematic
uncertainty associated with fitting the flux density of a faint source near a bright one, we performed 
a Monte-Carlo simulation. We identified a relatively isolated source in the mosaic
with about the same
flux density as the spiral galaxy: this is the galaxy at 3:32:43.49, -27:45:56.45 (J2000) 
which has a flux density of 121$\pm$10$\mu$Jy
at 16$\mu$m.
We added an artificial source at a distance between 7$\arcsec$ and 8$\arcsec$ from this 
source and extracted the flux density using the positional priors. The process was repeated 100 times
each for a range of flux densities straddling the measured flux density of the source. The 
extracted flux was compared with the input flux density to assess the systematic uncertainty and/or
flux bias in the measurement. For a source with a brightness comparable to JD2, the Monte-Carlo analysis yielded 1$\sigma$
uncertainties of 14$\mu$Jy, 12$\mu$Jy and 10$\mu$Jy at 16, 22 and 24\,\um, 
respectively. This
systematic uncertainty is primarily due to the extended wings of the brighter source and is a factor of
$\sim$2-3 larger than the statistical uncertainty. We therefore adopt values 
of 14$\pm$14$\mu$Jy at 16$\mu$m, 56$\pm$12$\mu$Jy at 22$\mu$m and 51$\pm$10$\mu$Jy at 24$\mu$m for JD2.

The sharp increase in flux density between 16$\mu$m and 22$\mu$m can either be due to the entrance of 
polycyclic aromatic hydrocarbons into the 22 and 24\um\ bandpasses or the 9.7$\mu$m silicate absorption feature entering the 
16$\mu$m bandpass. A third possibility is the decreasing contribution from hot dust surrounding an AGN
because the dust is close to its sublimation temperature. We evaluate each of these possibilities
in the following sections.

We note that the IRS is calibrated with respect to a constant $\nu$F$_{\nu}$ source spectrum. Color
correction terms to the observed flux are smaller than 1\% for the 16$\mu$m and 22$\mu$m data for the observed
source spectrum and can be neglected. Furthermore,
when we estimate fits to the observed photometry, we integrate template spectra over the filter bandpasses 
(Figure 2).

\section{Derived Source Parameters}

The optical/near-infrared photometry of JD2 has been discussed in \citet{Mobasher} and
\citet{Dunlop}. Spectral energy distribution fits to the photometry, which provide a photometric redshift
solution, show a bimodal distribution of 
minimum $\chi^{2}$ values at $z\sim2.5$ and $z\sim6.5$. One set of $\chi^{2}$ values are favored
over the other depending on the choice of optical photometry. \citet{Mobasher}, with their
stringent $BViz$ limits, derive the probability of a $z<5$ source to be 15\%. \citet{Dunlop} instead
 adopt marginal detections of the source in the $Viz$ bandpasses which we discuss in
Secion 3.2. Any significant detection
of the galaxy at optical wavelengths would rule out a photometric redshift solution which favors $z>6$.
\citet{Dunlop} find a minimum in the $\chi^{2}$ distribution at $z\sim2.2$.

The detection of HUDF-JD2 in the deep GOODS 24$\mu$m images did not break the degeneracy. 
\citet{Mobasher} suggests the hot dust emission is from an obscured AGN in the galaxy. 
Alternately, the 24\um\ emission could be dominated from polycyclic
aromatic hydrocarbon (PAH) features which
would be present if the object is an extremely red, dusty starburst galaxy at $z\sim2.5$ \citep{Yan:04, Stern:06}.  
Inclusion of the 16 and 22\um\ data allows us to measure a photometric redshift from the mid-infrared
which could potentially break this degeneracy,
independent of the
optical/near-infrared SED fits.
Although fitting optical to mid-infrared simultaneously is more elegant, it is virtually impossible
since there is virtually no correlation between optical/near-infrared SEDs and mid-infrared SEDs of galaxies.

\subsection{Mid-Infrared Photometric Redshifts}

The mid-infrared SED of a star-forming galaxy is a complicated interplay
of warm dust continuum, PAH emission and silicate absorption features. The
ratio of flux densities in the three abutting \sst\ passbands at 16, 22 and 24\um\ allow redshifts
of objects to be constrained as these dust features move through these filters. 
PAH are only 0.5-1$\mu$m wide, while the 9.7$\mu$m silicate feature is 3$-$4$\mu$m wide, depending on the continuum
level adopted. As a result, adjacent passbands can show widely different flux density ratios as a function
of redshift (Figure 3).

At $0.2<z<0.6$,
the 9.7\um\ silicate feature enters the 16$\mu$m bandpass while the 22 and 24\um\ bandpasses are
tracing the warm dust and 11 and 12\um\ PAH features. At these redshifts, sources
show low 16\um/22\um\ flux density ratios.
At $0.6<z<1.2$, the 9.7$\mu$m features moves out of the 16\um\ bandpass and into the 22\um\ bandpass,
while the 6.2\um\ and 7.7\um\ PAH features fall in the 16\um\ window, boosting the 16\um/22\um\ ratio.
At $1.2<z<1.7$, the 9.7\um\ feature moves into the 24\um\ band. Since the 22\um\ bandpass is $\sim$2\um\ wider
than the MIPS 24\um\ band, the 7.7\um\ feature enters the 22\um\ bandpass. The net effect is to boost the 
22\um/24\um\ flux ratio while lowering the 16\um/22\um\ ratio. At $z>1.9$, the PAH emission shifts out of the
16\um\ band, decreasing the 16\um/22\um\ ratio while the 22\um/24\um\ flux ratio varies as the 6.2 and 7.7\um\ PAH features
move within the passbands. This variation in flux ratios is illustrated in Figure 3 for four different SED
types derived from spectral observations of galaxies in the local Universe \citep{Armus, Brandl:04}.

While it is true that variation of PAH line ratios or strength of silicate absorption can cause broadband fluxes
to vary significantly, the combination of photometric redshifts from the optical/near-infrared and mid-infrared
photometric redshifts using these three bandpasses can help break redshift degeneracies.

The flux density ratios observed for HUDF-JD2 are a 16\um/22\um\ ratio of 0.25$^{+0.19}_{-0.08}$ and a 22\um/24\um\ ratio
of 1.1$^{+0.37}_{-0.13}$. This indicates that the galaxy must either be at $z\sim0.6$ if it has very strong silicate absorption or at
$z\sim1.7$ if it is a typical starburst galaxy with strong PAH emission. 

In Figure 4 and Table 1, we illustrate the quality of fits to the photometry at 16, 22 and 24$\mu$m from
different mid-infrared template spectral energy distributions. We consider starbursts, AGN and composite sources
which span the range of strong PAH, weak PAH and strong silicate 
absorption \citep{CE01, FS01, lef:01, Dale:02, Brandl:04, Armus}. 
We also consider the non-detection of JD2 in deep 70\um\ observations
of the GOODS fields which achieve 5$\sigma$ flux density limits of $\sim$2 mJy \citep{Frayer:06}.
The best fit to the 16$\mu$m,
22\,\um\ and 24\,$\mu$m photometry is from a starburst source with strong PAH emission at $z=1.7$. The increase
in the 22$\mu$m and 24$\mu$m flux relative to 16$\mu$m is due to the entrance of the 7.7$\mu$m PAH complex into the 22$\mu$m and
24$\mu$m bandpasses. 
The flux in the
16$\mu$m bandpass is lower since the 6.2$\mu$m PAH feature is much weaker. 

Extrapolating the templates that are fit to the mid-infrared photometry also yields an estimate of the far-infrared
luminosity for the galaxy.
The L$_{\rm IR}$=L(8$-$1000\um) luminosity of the source is $\sim$5$\times$10$^{11}$~L$_{\sun}$,
which corresponds to a star-formation rate of 80~M$_{\sun}$ yr$^{-1}$. 
We compare this with optical/UV estimates of star-formation in the following section.
Using the radio-FIR correlation in \citet{Yun}, we predict the 1.4 GHz to be 12$\mu$Jy and the 8.4 GHz flux density to be
3.7$\mu$Jy while the predicted fluxes at 70\um\ and 850\um\ are shown in Table 1. The deepest 70\um\ and 850\um\ surveys
are currently sensitive to 1.5$-$2 mJy. The
predicted intensities of JD2 are below these limits implying that confirmation of the derived
far-infrared luminosity will have to await deeper observations by {\it Herschel} and ALMA.

$\chi^{2}$ values are also reduced for an obscured AGN template at $z=0.6$ which has
the observed
16\um\ flux density suppressed relative to the flux density at 22\um\ and 24\um\ due to the entrance of the 9.7\,\um\
silicate absorption feature. 
An obscured AGN template like Mrk231 can also fit the 8\um\ flux density of the source as shown in Figure 4
while the starburst template fits at $z\sim1.7$ require the optical/near-infrared flux to be dominated
by starlight (Figure 6). The non-detection
of any spectral lines in the optical/near-infrared spectroscopic
data presented by \citet{Mobasher}, as well as the difficulty in accounting for the non-detection of the source
at optical wavelengths (Section 3.3)
makes the $z\sim0.6$ hypothesis unlikely. On the other hand, $z\sim1.7$ lies within the so-called
redshift ``desert", where the [OII] $\lambda$3727 doublet, 
[OIII] $\lambda$5007 line and H$\alpha$+[NII] lines are all severely affected
by atmospheric transmission and strong OH sky lines. Furthermore, as we discuss in Section 3.3, at $z=1.7$, the red
SED of the source between 2.2$\mu$m and IRAC 3$-$8\um\ could be due to the broad 1.6\um\ bump
in the stellar SED being in the IRAC passbands.

We now consider the implications of the IRS detections to the high redshift advocated in \citet{Mobasher}.
If we adopt the hypothesis that JD2 is at $z\sim6.5$ and the 24\um\ emission is dominated by AGN light, we 
must first subtract the stellar contribution to
the mid-infrared flux densities. We find that for the SED fit by \citet{Mobasher}, the contribution of stellar
photospheric emission to the flux densities at 16, 22 and 24\um\ are 10.7, 6.5 and 6.0 $\mu$Jy respectively. 
We then attempt to fit the starlight-subtracted mid-infrared photometry with the two mid-infrared AGN
templates presented in \citet{Mobasher}.
The $\chi^{2}$ values are significantly worse than the best fits at $z\sim1.7$ (Table 1 and Figure 5). The Mrk231 template, fairly
typical of an obscured AGN, is bluer in its 16\um/22\um\ flux ratios at $z=6.5$ than what the observations indicate.
NGC1068, which is a Compton-thick AGN with decreasing hot dust continuum at shorter wavelengths due to dust sublimation, 
would best fit the observed mid-infrared photometry at $z=3.7$. The hot dust emission falls off too rapidly for it to
account for the 22\um\ flux density with respect to the 24\um\
flux density if it were at $z=6.5$. One could imagine that varying the AGN
template could allow for the mid-infrared photometry here to be fit but given that the $z=1.7$ fit using typical
mid-infrared templates agrees with the parameters derived from the optical/near-infrared fits, we find the
$z=6.5$ interpretation for JD2 substantially weakened.

\subsection{The Optical/Near-infrared Photometry}

Given the derived redshift based on the mid-infrared photometry of the source, we critically investigate the
optical/near-infrared photometry to assess potential sources of error.

The ACS $BViz$ photometric measurements are very important to the interpretation
of JD2, since a significant optical detection would almost certainly rule out the
$z \approx 6.5$ hypothesis.   We note that \citet{Mobasher} erroneously
reported that their ACS photometry was measured in an $0\farcs9$ diameter
aperture, when in fact a smaller, $0\farcs48$ diameter aperture was used
(M.\ Dickinson, private communication).   A larger aperture size would lead to more
conservative photometric limits, and is probably to be recommended, given the
size of the galaxy as measured from the NICMOS $H$-band images
(measured half-light radius $0\farcs3$, uncorrected for PSF effects, see Mobasher et al.).
However, a larger aperture risks including light from three faint, neighboring galaxies
located about 1\arcsec\ away from JD2.    We have masked out regions with
diameter $0\farcs9$ around those three galaxies\footnote{In another typographical
error, \citet{Mobasher} reported masking regions $0\farcs5$ in diameter around
the neighboring galaxies.  In fact, the masking diameter was $0\farcs9$, which we
also adopt here.} before measuring photometry for JD2 in a $1\farcs0$ diameter circular aperture.

The ACS images were drizzled using a point kernel \citep[see][]{Beckwith} which
should lead to noise that is uncorrelated between adjacent pixels.  We verified that this
is the case by measuring the autocorrelation function of the noise after masking galaxies.
We measured the noise on various scales, including 50 pixel apertures as used by
\citet{Beckwith}.  Our most conservative (largest) noise
measurements are 31 to 37\% smaller than those reported by \citet[][their Table 5]{Beckwith}.

\citet{Dunlop} report faint detections of positive flux in the ACS $V$, $i$
and $z$-bands, which, taken together, drive their photometric redshift estimate
to favor a lower value, $z \approx 2.15$.   We do not reproduce these measurements;
using a $1\farcs0$ diameter aperture and our noise measurements, and the 
correction of $\sim$0.1 mag for the energy falling outside the aperture
reported in \citet[][Table 3]{Sirianni}, we find
the $2\sigma$ photometric limits given in Table~2.  These are
$\sim$0.8 mag brighter than the limits reported in Mobasher et al.\ (2005), mainly
due to the larger aperture adopted here.  It is possible that the positive flux detected
in the Dunlop et al.\ measurements arises in part from the faint neighboring galaxies. 

The NICMOS $J_{110}$ photometry and $H_{160}$ photometry in \citet{Mobasher} are very similar to those in the UDF
catalog of \citet{RIT}. As a result, we have not remeasured these values.
However, we note that the photometric values in \citet{Mobasher} do not include the NICMOS count rate dependent non-linearity
correction discussed in \citet{deJong}. The effect of this non-linearity correction is to make the photometry
brighter by 0.22 mag in the $J_{110}$ band and 0.12 mag in the $H_{160}$ band as shown in Table 2. 

We also remeasured the photometry of the source at IRAC wavelengths. We used SExtractor catalogs
and measured the photometry in 3$\arcsec$ diameter beams, similar to \citet{Mobasher}. To derive
aperture corrections and systematic uncertainties, we adopt a different approach. The primary
systematic uncertainty in this measurement is the uncertain sky level due to the bright galaxy
7$\arcsec$ to the South-East. We input an artificial point source whose brightness is that of JD2 into
the final mosaic.
The centroid of the source with respect to the bright galaxy is kept similar to that of JD2,
i.e the artificial source has a centroid which is between
7$\arcsec$ and 8$\arcsec$ from the galaxy. The SExtractor routine was
run on this fake image and the photometry of the fake source measured. We find a systematic uncertainty
whereby the extracted flux of the fake source in the catalog was brighter at 3.6 and 4.5\um\ but
fainter at 5.8 and 8.0\um. 
The process was repeated a 100 times to measure the average systematic and statistical error
using the corresponding point spread function at each of the four IRAC wavelengths.
We revise the aperture corrected magnitudes for JD2 to 22.24, 22.00, 21.63 and 21.51 AB mag at
3.6, 4.5, 5.8 and 8.0 \um\ respectively. There is, however, a correction to be applied
to these magnitudes.
 
There are three faint galaxies within 1$\arcsec$
of JD2. Two of these are rather faint and blue while the brightest of the three falls within the IRAC 3$\arcsec$ beam.
These galaxies are detected at all wavelength between the $B$ and $H_{160}$ band. 
To estimate the contribution of these sources to the photometry within the IRAC beam, we fit \citet{BC03} templates
to the multiband photometry of these sources, leaving redshift as a free parameter. We use the best fit SEDs
to estimate the redshift and corresponding flux densities at the IRAC wavelengths. We find the brightest 
of these three sources, located at 3:32:38.76,-27:48:28.91 (J2000), to be at redshift 3.4 with magnitudes
of 25.95, 25.83, 25.82 and 25.80 at 3.6, 4.5, 5.8 and 8.0\um\ respectively. The second source, located
at 3:32:38.81,-27:48:39.79, is at $z\sim3.1$ with a contribution that is about 0.6 mag fainter at each of the
four passbands. We subtracted the contribution of both these sources from the photometry of JD2.

To summarize,
the corrected limits for JD2 in the ACS bands are about 0.8 mag brighter
due to the larger aperture, the photometry in the
NICMOS $J_{110}$ and $H_{160}$ bands is brighter by $\sim10-20$\% due to the non-linearity 
correction and the photometry in the IRAC bands is about 10\% fainter on average than those
adopted in \citet{Mobasher}. The net effect is to reduce the amplitude of the apparent near-infrared
to IRAC ``break" slightly.
The optical/near-infrared photometry of JD2 after these corrections have been applied
is shown in Table 2 and which we fit for in Section 3.3.  For the sake of 
completeness, we also investigate redshift constrained fits to the \citet{Dunlop} and \citet{Mobasher}
optical/near-infrared photometry.

In addition, we attempted to fit elliptical isophotes to the galaxy to the SE and subtracted it out from the image.
This process left significant residuals due to the asymmetric nature of the \sst/IRAC point spread function. 
The aperture corrected point source photometry we measure for JD2 in images with the galaxy subtracted
is 22.36, 21.95, 21.72 and 21.64 AB mag. The difference between these values and those
quoted in Table 2 can be attributed
almost entirely to the difficulty in measuring the absolute sky background. Thus, we provide photometric
uncertainties of 0.15 mag for all 4 bands.

\subsection{Revisiting the Optical/Near-Infrared Photometric Fits}

We first evaluate the quality of the fits to the optical/near-infrared photometry at $z=0.6$ using
the \citet{BC03} population synthesis models. The range of parameter space for the models was
solar and 0.2 solar metallicity, 19 e-folding timescales ($\tau$) for the starburst from
an instantaneous starburst to constant star-formation, ages ($t$) from 0.1 Myr to the age of the Universe
at the chosen redshift and extinction (A$_{\rm V}$) between 0 and 10 mags.
At $z=0.6$,
the 8\um\ flux density is affected by dust emission and is ignored in the fits.

If JD2 is a $z=0.6$ starburst like NGC6240 but with silicate absorption which suppresses the observed
16\um\ flux density, the weakness/non-detection of the source in the $BViz$ passbands requires A$_{\rm V}$=9.9 mag
of visual extinction irrespective of whether we adopt the \citet{Dunlop} or \citet{Mobasher} photometry.
This is not impossible but unusually large. It would imply a young 8 Myr old starburst with a mass of
2$\times$10$^{9}$~M$_{\sun}$ and a true star-formation rate, as derived
from extinction correction of the ultraviolet flux, of $\sim$100~M$_{\sun}$~yr$^{-1}$, all of which
is dust obscured. The fits to the mid-infrared photometry at $z=0.6$ imply a much lower dust-obscured star-formation rate of
$\sim$5~M$_{\sun}$~yr$^{-1}$. Although it is possible for optically thick star-formation, such as that found in local
ULIRGs, to result in the mid- and far-infrared luminosity of galaxies exceeding their extinction corrected ultraviolet
luminosity, the converse is rarely true.  This weakens the possibility of a $z=0.6$ starburst. 

If the source is a $z=0.6$ obscured AGN like Mrk231, then the contribution of the hot dust
around the AGN to the near-infrared photometry needs to be subtracted before fitting. For the Mrk231 template,
we find this to be 0.8, 1.2, 
1.4, 2.6, 3.5, 5.6 and 10.5$\mu$Jy
at 1.2, 1.6, 2.2, 3.6, 4.5, 5.8 and 8.0$\mu$m respectively. The photometry at 1.1 and 1.6\um\ of the source is
a factor of $3-10$ below the expected values again indicating that a Mrk231 type source, while consistent
with the mid-infrared photometry, cannot fit the entire SED.

Thus, a $z=1.7$ dusty star-forming galaxy SED is most consistent with all the photometry. We adopt this redshift and
re-fit the optical/near-infrared data using the BC03 models. Results are shown in Table 3. 
HUDF-JD2 is best fit with a solar metallicity template from the BC03 library with A$_{\rm V}\sim4$~mag
of visual extinction using a \citet{Calzetti:01} dust extinction law. We caution that
the metallicity is not strongly constrained. The best-fit model is a post-starburst BC03 SED
with $t=570$~Myr, $\tau$=30 Myr and a stellar mass of 6$\times$10$^{10}$~M$_{\sun}$.
This is similar to the $z\sim2.5$ fits presented in \citet{Mobasher}. 
The result is not surprising since post-starburst galaxies
could have significant dust content from dust produced in supernovae and AGB stars. In fact,
it has been observationally shown that the host galaxies of core-collapse and Type Ia supernovae are
dustier than field galaxies with the same observed optical brightness \citep{Chary05}. 
We caution however, that the post-starburst hypothesis, motivated by the large value of
$t/\tau$, would be weakened if most of the star-formation
takes place in optically thick regions of the galaxy.

From the SED fits, we estimate that the star-formation rate, as measured by
the 1500\AA\ flux escaping the galaxy, is very small ($\sim$2$\times$10$^{-4}$~M$_{\sun}$~yr$^{-1}$).
Application of an extinction
correction to the UV luminosity results in a star-formation rate of $<$1~M$_{\sun}$~yr$^{-1}$. 
In contrast, its true star-formation rate as
derived from the mid-infrared fits is $\sim$80-90~M$_{\sun}$~yr$^{-1}$. The 
fact that JD2 is a compact near-infrared source with a scale length of $\sim$0.3$\arcsec$, combined
with the large ratio of infrared to ultraviolet star-formation rate estimates, strongly suggests that
JD2 is being powered by a compact, optically-thick starburst in its nucleus.

There is a caveat. If we use the \citet{BC03} models to solve for an independent redshift from the 
optical/near-infrared photometry presented
in this paper, we continue to get a high redshift minimum at $z=7.2$ which has a reduced $\chi^{2}$ of
1.05 compared to the reduced $\chi^{2}$ values of 1.9 and 2.4 at $z=2.4$ and $z=1.7$ respectively.
Thus, optical/near-infrared photometric redshifts continue to favor the high redshift hypothesis,
while mid-infrared photometric redshifts favor the lower redshift hypothesis. We favor the mid-infrared
photometric redshifts derived here since optical photometric redshifts for dusty galaxies have been shown to have a
large scatter based on a comparison between spectroscopic and photometric redshifts for 24\um\ detected sources
in GOODS-N \citep{Chary:06}.

A comparison of the properties of HUDF-JD2 with the massive, red galaxy population at $1.5<z<3$ is illustrative.
\citet{Papovich} show that the star-formation rate of galaxies scale with the stellar mass at these redshifts
and that the specific star-formation rate of a dusty starburst with mass of $\sim$6$\times$10$^{10}$~M$_{\sun}$ is
about $\sim1-20$ Gyr$^{-1}$. UDF-JD2, for $z\sim1.7$, has a specific star-formation 
rate of 1.6 Gyr$^{-1}$ which is consistent with the massive, dusty starbursts at these redshifts. The optical extinction
derived for this galaxy is rather large, corresponding to $E(B-V)$=1.2. Spectroscopically confirmed galaxies
at these redshifts which are 24$\mu$m detected have $E(B-V)$ values derived from their ultraviolet
slopes of $\sim$0.3. This is most likely a bias since spectroscopic confirmation requires the detection of
optical/ultraviolet lines which are preferentially detected in galaxies that are relatively transparent 
and have low extinction values. 
A better comparison is with the derived extinction values for local dusty starbursts
presented in \citet{Hopkins}. The average $E(B-V)$ to the stellar continuum
for that sample at an infrared luminosity of 5$\times$10$^{11}$~L$_{\sun}$
is about 0.25. Thus, HUDF-JD2 does not appear to be dissimilar from dusty, luminous infrared
galaxies (LIRGs) at $z\sim2$. However, its
derived $E(B-V)$ appears to be a factor of $\sim$4 higher than other LIRG samples, consistent with
its significant detection at mid-infrared wavelengths and compact morphology in the NICMOS 1.6\,\um\ image which
suggests a compact, nuclear starburst similar to many local infrared luminous galaxies.

JD2 is about 1.4 mags redder in observed $R-$[3.6] colors and about 3 mags fainter
at 3.6\,\um\ than HR10, a $z=1.44$ spectroscopically confirmed dusty starburst \citep{Stern:06}. 
The predicted H$\alpha$ line flux for this galaxy, including the extinction in the line,
is $\sim$10$^{-18}$\,erg\,\,s$^{-1}$~cm$^{-2}$. This is a factor of 3 below
the sensitivity of the existing observations in \citet{Mobasher} but should be detectable with a space-based
near-infrared spectrograph.

\section{Conclusions}

We present new 16 and 22\um\ mid-infrared photometry of the HUDF source JD2 which had tentatively been classified
as a $z\sim6.5$ massive galaxy. The mid-infrared spectral energy distribution shows a sharp increase in flux
density between 16\um\ and 22\um, most consistent with the 7.7\um\ PAH feature entering the 22\um\ bandpass.
Our analysis concludes that the galaxy is a starbursting, luminous infrared galaxy (LIRG) at a redshift of $z\sim1.7$.
The non-detection of the source 
in the UDF $BViz$ data is due to the presence of $A_{\rm V}\sim4$ mag of dust internal to the galaxy.
We have re-evaluated the optical/near-infrared photometry of this source and find the stringent
limits in \citet{Mobasher} are more consistent with the data rather than the marginal detection reported in
\citet{Dunlop}. Photometric redshift fits to the optical/near-infrared data alone continue to favor a $z>6$ solution.
However, the unusually high L$_{\rm MIR}$/L$_{X}$ for JD2 compared to other obscured AGN, the large
scatter in optical photometric redshift solutions for dusty galaxies and the similarity
between the multiwavelength properties of this galaxy and other dusty starbursts in the $1.5<z<3$ range lead
us to favor the mid-infrared photometric redshift solution.
The non-detection of a $z\sim2$ LIRG in the deepest optical data taken to date
strongly cautions against interpreting near-infrared dropout sources in shallower surveys as 
$z>6$ galaxies. 

\acknowledgements 
We wish to thank Dave Frayer and Megan Eckart for useful advice.
This work is based on observations made with the {\it Spitzer} Space
Telescope, which is operated by the Jet Propulsion Laboratory, California
Institute of Technology, under a NASA contract. Support for this work
was provided by NASA through an award issued by JPL/Caltech.

`
\begin{figure}
\epsscale{0.7}
\plotone{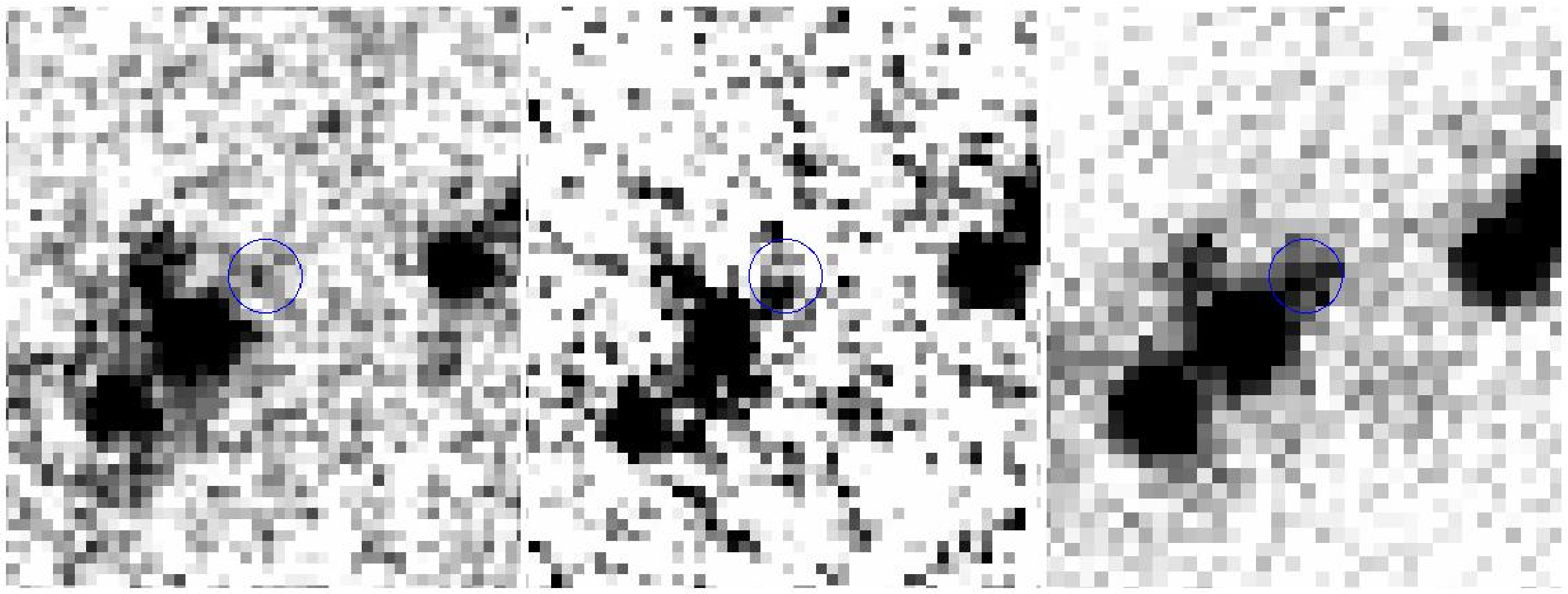}
\caption{
16, 22 and 24\um\ snapshots of the \hst\ Ultradeep Field source JD2 (circled) illustrating the quality of
\sst\ mid-infrared data on this source. Images are $\sim$40$\arcsec$ on a side with North up and East to the left.
JD2 is at 3:32:38.74, -27:48:39.9 (J2000).
}
\end{figure}

\begin{figure}
\plotone{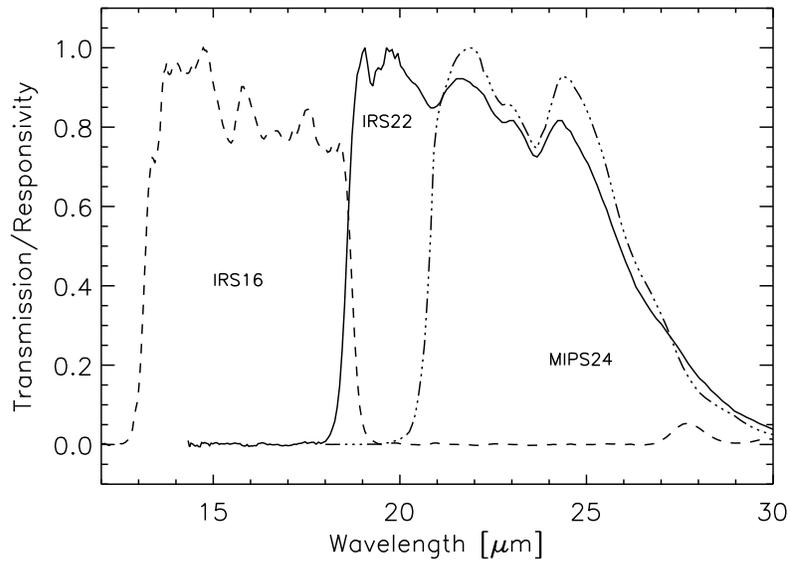}
\caption{Response curves for the IRS 16\um\ and 22\um\ bandpasses and MIPS 24\um\ bandpass.}
\end{figure}

\begin{figure}
\plotone{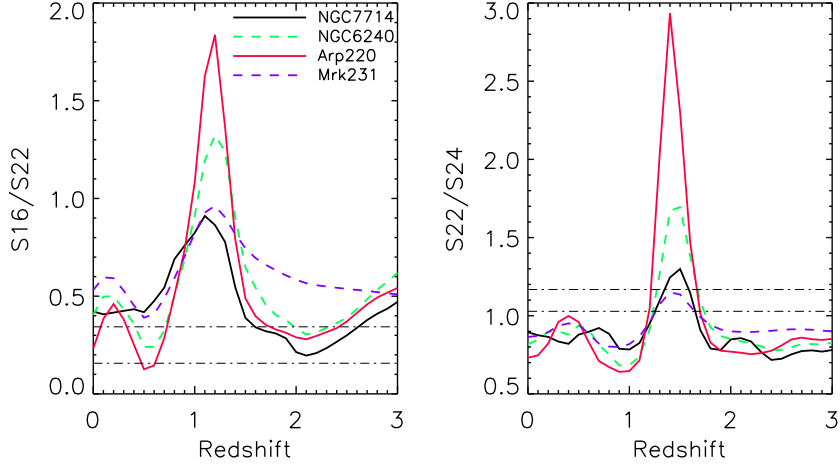}
\caption{
16\um/22\um\ and 22\um/24\um\ flux density ratios as a function of redshift
for typical infrared luminous galaxies at 0$<z<3$. 
The sources shown are NGC7714 - a typical starburst (SB) with strong PAH, NGC6240 - a composite AGN+SB object with
strong PAH, Arp220 - a starburst with strong silicate absorption and weak PAH
and Mrk231 - an AGN with weak silicate absorption. The range of flux
ratios of HUDF-JD2 with statistical errors is shown by the horizontal dot-dash lines. 
The flux ratios indicate a starburst with strong PAH at $z\sim1.7$ or a source with strong silicate absorption at $z\sim0.6$.
}
\end{figure}

\begin{figure}
\plotone{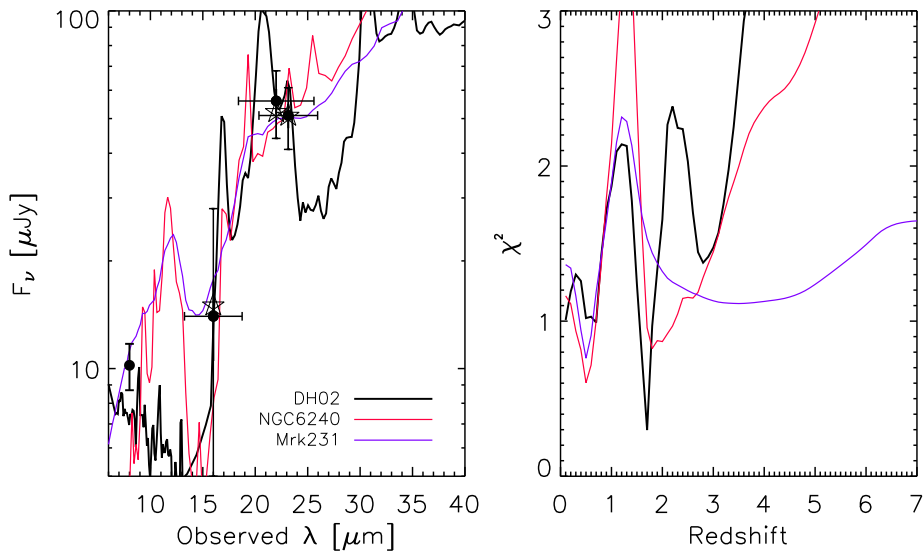}
\caption{
(Left panel) Spectral energy distribution fits to the mid-infrared
photometry of HUDF-JD2 along with the distribution of $\chi^{2}$ values
with redshift (Right panel). The SEDs are plotted at their best fitting redshift
which is $z=1.7$ for the \citet{Dale:02} starburst templates, $z=0.5$ for NGC6240 and $z=0.5$ for Mrk231. 
The preferred fit (i.e. lowest $\chi^{2}$ value) indicates the source is a luminous infrared
galaxy with strong PAH emission at $z\sim1.7$. A low $\chi^{2}$ is also obtained
at $z\sim0.6$ from a source with either strong silicate absorption or the gap between the
8 and 11 $\mu$m PAH complexes. However, the $z\sim1.7$
fit is more consistent with the fits to the optical/near-infrared
photometry.  The stars in the left panel are the \citet{Dale:02} SED convolved
with the filter curves.
}
\end{figure} 

\begin{figure}
\plotone{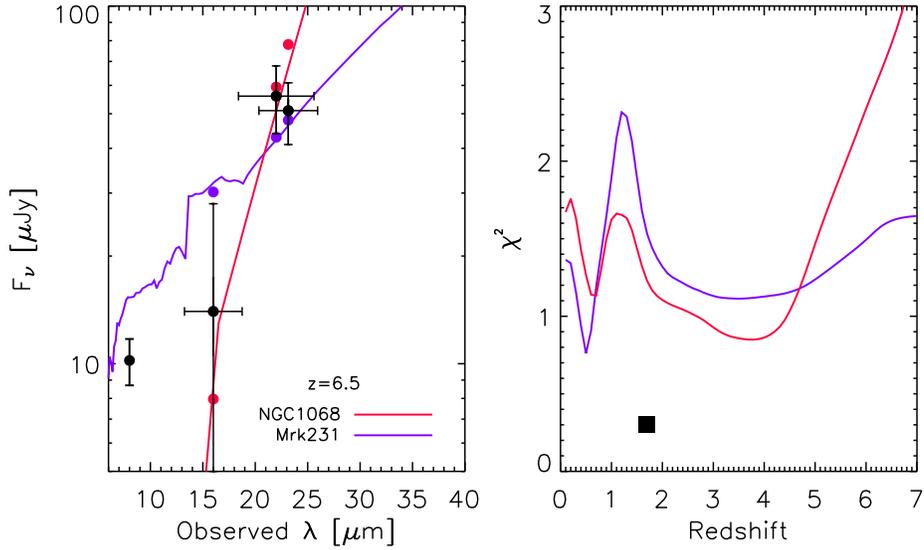}
\caption{
(Left panel) Fits to the mid-infrared
photometry of HUDF-JD2 at $z=6.5$ using the two AGN templates presented in \citet{Mobasher}.
The solid red and purple circles are the AGN templates of the corresponding color
integrated through the relevant bandpasses.
The right panel shows the distribution of $\chi^{2}$ values. The solid black square shows
the minimum $\chi^{2}$ obtained at $z=1.7$ for the starburst spectral energy distribution
discussed in Figure 4 and in Section 3.1. If the source were really at $z=6.5$, there
would be a stellar contribution to the mid-infrared photometry as discussed in the text.
Although, this improves the formal $\chi^{2}$ estimates, the $z=6.5$ fits still yield substantially worse $\chi^{2}$
estimates than the best fit at $z=1.7$ as shown in Table 1.
}
\end{figure}

\begin{figure}
\plotone{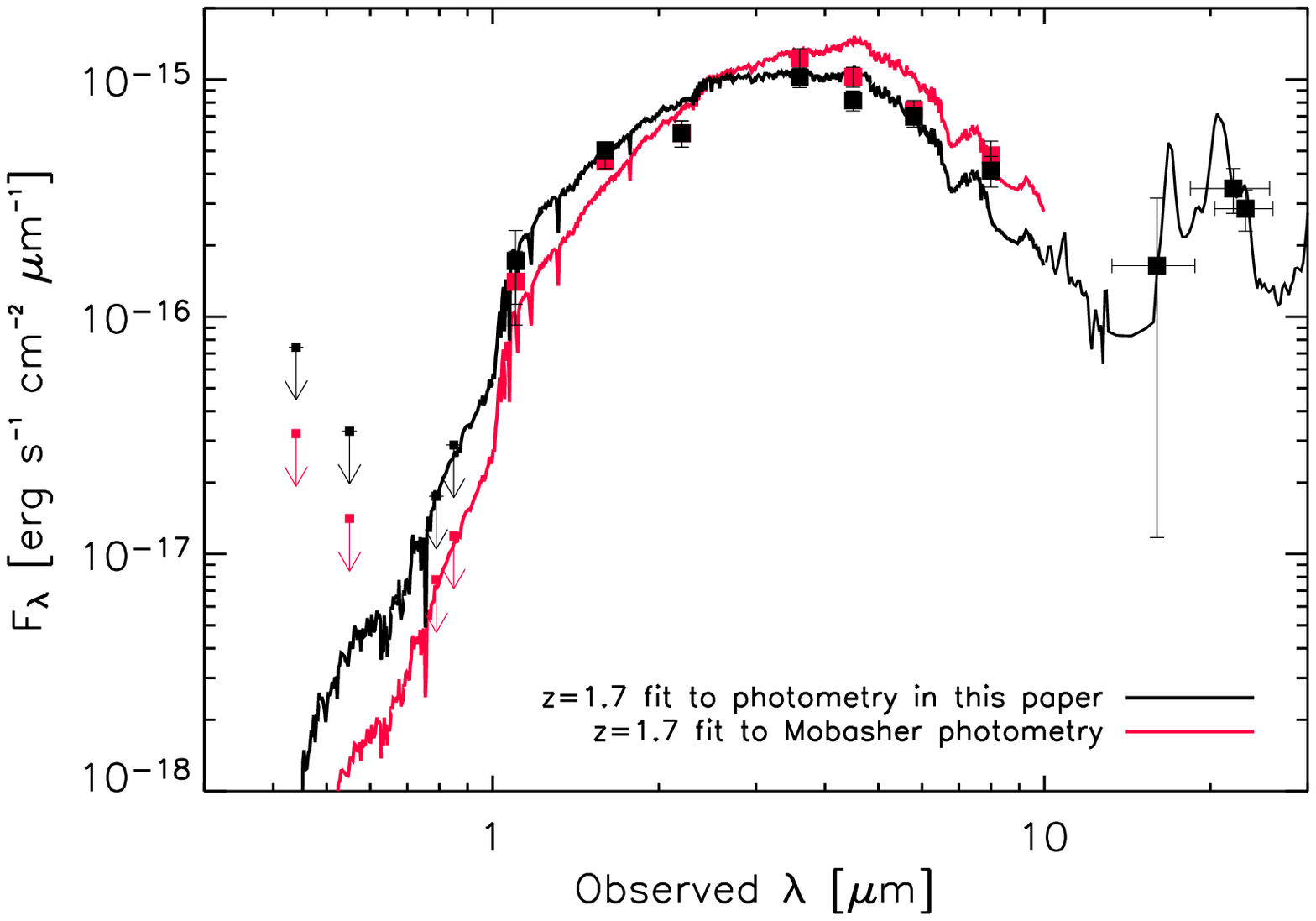}
\caption{
Optical to mid-infrared spectral energy distribution of HUDF-JD2. Solid black squares are the 
photometry presented in this paper. Solid red squares are the photometry in \citet{Mobasher}. 
The optical/near-infrared
photometry has been fit with the \citet{BC03} model and the mid-infrared photometry fit
with the \citet{Dale:02} model.
The source is most likely a luminous infrared galaxy (LIRG) with strong PAH emission at $z\sim1.7$. 
The inferred star-formation rate is $\sim$80~M$_{\sun}$~yr$^{-1}$ with $\sim$4 mag of visual
extinction.
}
\end{figure}

\begin{deluxetable}{lcccccl}
\tabletypesize{\scriptsize}
\tablecaption{Results of Mid-infrared Spectral Energy Distribution Fits}
\tablewidth{0pt}

\tablehead{
\colhead{Template Adopted} &
\colhead{Best fit redshift} &
\colhead{L$_{\rm IR}$\tablenotemark{a}} &
\colhead{$\chi^{2}$} &
\colhead{Predicted Fluxes} &
\colhead{Type of Object\tablenotemark{b}} &
\colhead{Notes} \\

\colhead{} &
\colhead{} &
\colhead{L$_{\sun}$} &
\colhead{} &
\colhead{(70, 850$\mu$m in mJy)} &
\colhead{} &
\colhead{}
}
\startdata
Dale \& Helou 	&	1.7	&	4.7$\times$10$^{11}$&	  0.30	&	0.6, 0.5&		   ... &\\
Chary \& Elbaz	&	1.6	&	5.6$\times$10$^{11}$ &   0.37   &	0.9, 0.4&		   ... &\\
NGC7714 &	        1.6     &       2.5$\times$10$^{11}$ &   0.42   &	0.4, 0.2 & 		Starburst&\\
M82	&		1.7	&	1.6$\times$10$^{11}$&	  0.48	&	0.2, 0.09&	   Starburst &\\
NGC6240	&		0.5	&	1.5$\times$10$^{10}$&    0.60	&	1.0, 0.02&	   Compton thick AGN+SB & large A$_{\rm V}$ \\ 
&&&&&&inconsistent UV/IR SFR\\
Mrk231	&		0.5	&	6.9$\times$10$^{9}$&	  0.76	&	0.4, 0.004&	   Obscured AGN & exceed NIR flux\\
Arp220	&		1.8	&	2.1$\times$10$^{12}$&	  0.95	&	1.7, 1.5&	   Starburst & \\
NGC1068-nucleus &		3.7	&	1.4$\times$10$^{12}$&	  1.02	&	0.3, 1.0&	   Obscured AGN & Worse $\chi^{2}$\\
&&&&&&\\
Mrk231	&		6.5	&	7.4$\times$10$^{12}$&	  2.0	&	0.2, 2.4&	   Obscured AGN & Worse $\chi^2$ \\
NGC1068-nucleus &		6.5	&	1.3$\times$10$^{13}$&	  1.1	&	0.6, 7.1&	   Obscured AGN & Worse $\chi^2$ \\

\enddata
\tablenotetext{a}{For the star-forming templates, the star-formation rate is 1.71$\times$10$^{-10}\times$L$_{\rm IR}$.}
\tablenotetext{b}{For the AGN templates at $z=6.5$, the stellar photospheric emission has been subtracted from the mid-infrared photometry. To identify the best-fit mid-infrared redshift for the AGN templates the actual mid-infrared photometry was used.}
\end{deluxetable}

\begin{deluxetable}{ll}
\tabletypesize{\scriptsize}
\tablecaption{Revised Optical/Near-infrared Photometry for JD2}
\tablewidth{0pt}

\tablehead{
\colhead{Wavelength} &
\colhead{Photometry (AB mag)\tablenotemark{a}}
}
\startdata
{\it B} & $>$29.7 \\
{\it V} & $>$30.1 \\
{\it i} & $>$30.0 \\
{\it z} & $>$29.3 \\
$J_{110}$ & 26.8$\pm$0.3 \\
$H_{160}$ & 24.82$\pm$0.07 \\
$K_s$ & 23.95$\pm$0.132 \\
3.6\um\ & 22.29$\pm$0.15 \\
4.5\um\ & 22.05$\pm$0.15\\
5.8\um\ & 21.67$\pm$0.15\\
8.0\um\ & 21.54$\pm$0.15\\
16\um\ & 21.04$_{-0.76}^{+2.86}$ \\
22\um\ & 19.53$\pm$0.21 \\
24\um\ & 19.63$\pm$0.21 \\
\enddata
\tablenotetext{a}{Photometric limits, where provided, are 2$\sigma$. See text for details.}
\end{deluxetable}

\begin{deluxetable}{lccccccl}
\tabletypesize{\scriptsize}
\tablecaption{z=1.7 SED Fits to Optical-Near-infrared Photometry}
\tablewidth{0pt}

\tablehead{
\colhead{Mass} &
\colhead{A$_{\rm V}$} &
\colhead{Age} &
\colhead{$\tau$} &
\colhead{UV SFR} &
\colhead{Corrected SFR} &
\colhead{$\chi^{2}$\tablenotemark{a}} &
\colhead{Notes} \\

\colhead{M$_{\sun}$} &
\colhead{mag} &
\colhead{Myr} &
\colhead{Myr} &
\colhead{M$_{\sun}$~yr$^{-1}$} &
\colhead{M$_{\sun}$~yr$^{-1}$} &
\colhead{} &
\colhead{}
}
\startdata
{\bf 6.3$\times$10$^{10}$} & {\bf 3.6} & {\bf 570} & {\bf 30}  & {\bf 1.7$\times$10$^{-4}$} & {\bf 0.74} & {\bf 2.4} & {\bf Photometry from Table 2, Solar metallicity BC03} \\
9.8$\times$10$^{10}$ & 4.7 & 400  & 0   & 8$\times$10$^{-5}$ & 5 & 3.9 & Mobasher et al. photometry, Solar metallicity BC03 \\
1.3$\times$10$^{11}$ & 4.6 & 1020 & 30  & 4$\times$10$^{-5}$ & 1.8 & 4.0 & Mobasher et al. photometry, 0.2 solar metallicity BC03 \\
4.6$\times$10$^{10}$ & 4.3 & 453  & 700 & 0.003 & 93  & 1.6 & Dunlop et al. photometry, Solar metallicity BC03 \\
1.9$\times$10$^{10}$ & 4.7 & 47.5 & 50  & 0.005 & 308 & 1.6 & Dunlop et al. photometry, 0.2 solar metallicity BC03 \\
\enddata
\tablenotetext{a}{4 degrees of freedom with 7 data points and 3 parameters (Age, Extinction and $\tau$) for the \citet{Mobasher} 
photometry. The \citet{Dunlop} photometry has 10 data points.}
\end{deluxetable}


\begin{thebibliography}{}
\bibitem[Armus et al.(2007)]{Armus}
Armus, L., et al., 2007, ApJ, in press, astro-ph/0610218

\bibitem[Beckwith et al.(2006)]{Beckwith}
Beckwith, S. V. W., et al., 2006, AJ, 132, 1729

\bibitem[Brandl et al.(2004)]{Brandl:04}
Brandl, B., et al., 2004, ApJS, 154, 188

\bibitem[Bruzual \& Charlot\,(2003)]{BC03}
Bruzual, G., \& Charlot, S., 2003, MNRAS, 344, 1000

\bibitem[Calzetti\,(2001)]{Calzetti:01}
Calzetti, D., 2001, PASP, 113, 1449

\bibitem[Chary\,(2006)]{Chary:06}
Chary, R., 2006, Proceedings of ``At the Edge of the Universe'', astro-ph/0612736

\bibitem[Chary et al.(2005)]{Chary05}
Chary, R., et al., 2005, ApJ, 635, 1022

\bibitem[Chary \& Elbaz\,(2001)]{CE01}
Chary, R., \& Elbaz, D., 2001, ApJ, 556, 562

\bibitem[Chen \& Marzke\,(2004)]{Chen}
Chen, H.-W., \& Marzke, R. O., 2004, ApJ, 615, 603

\bibitem[Dale \& Helou\,(2002)]{Dale:02}
Dale, D., \& Helou, G., 2002, ApJ, 576, 159

\bibitem[Dav\'{e} et al.(2006)]{Dave:06}
Dav\'{e}, R., Finlator, K., \& Oppenheimer, B. D., 2006, MNRAS, 370, 273

\bibitem[de Jong et al.(2006)]{deJong}
de Jong, R., et al., 2006, Proceedings of The 2005 HST Calibration Workshop, eds. A. Koekemoer, P. Goudfrooij, L. Dressel, 121

\bibitem[Dickinson et al.(2003)]{MED03}
Dickinson, M., et al., 2003, Proceedings of the Mass of Galaxies at Low and High Redshift, 324, Springer-Verlag

\bibitem[Dunlop et al.(2007)]{Dunlop}
Dunlop, J., Cirasuolo, M., \& McLure, R. J., 2007, MNRAS, submitted, astro-ph/0606192

\bibitem[Forster-Schreiber et al.(2001)]{FS01}
Forster-Schreiber, N. M., et al., ApJ, 552, 544

\bibitem[Frayer et al.(2006)]{Frayer:06}
Frayer, D., et al., 2006, ApJ, 647, L9

\bibitem[Fruchter \& Hook(2002)]{Fruchter}
Fruchter, A. S., \& Hook, R. N., 2002, PASP, 114, 144

\bibitem[Hopkins et al.(2001)]{Hopkins}
Hopkins, A. M., et al., 2001, AJ, 122, 288

\bibitem[Houck et al.(2004)]{Houck:04}
Houck, J., et al., 2004, ApJS, 154, 18

\bibitem[Krabbe et al.(2001)]{Krabbe}
Krabbe, A., Boeker, T., \& Maiolino, R., 2001, ApJ, 557, 626

\bibitem[Le Floc'h et al.(2001)]{lef:01}
Le Floc'h, E., et al., 2001, A\&A, 367, 487

\bibitem[Mobasher et al.(2005)]{Mobasher}
Mobasher, B., et al., 2005, ApJ, 635, 832

\bibitem[Papovich et al.(2006)]{Papovich}
Papovich, C., et al., 2006, ApJ, 640, 92

\bibitem[Sirianni et al.(2005)]{Sirianni}
Sirianni, M., et al., 2005, PASP, 117, 1049

\bibitem[Stern et al.(2006)]{Stern:06}
Stern, D., Chary, R., Eisenhardt, P., \& Moustakas, L., 2006, AJ, 132, 1405

\bibitem[Thompson et al.(2005)]{RIT}
Thompson, R. I., et al., 2005, AJ, 130, 1

\bibitem[Teplitz et al.(2006)]{Teplitz:07}
Teplitz, H. I., et al., 2006, BAAS, 2091, 3203 

\bibitem[Yan et al.(2004)]{Yan:04}
Yan, H., et al., 2004, ApJ, 616, 63

\bibitem[Yun et al.(2001)]{Yun}
Yun, M. S., Reddy, N. A., \& Condon, J., 2001, ApJ, 554, 803

\end{thebibliography}
\end{document}